# Wavelength scaling of high-order harmonic yield from an optically prepared excited state atom

J. Chen[1, 3 §], Ya Cheng[2, †], and Zhizhan Xu[2, ‡]

[1] Institute of Applied Physics and Computational Mathematics,

P.O. Box 8009 (28), Beijing 100088, China

[2] State Key Laboratory of High Field Laser Physics, Shanghai Institute of Optics and

Fine Mechanics, Chinese Academy of Sciences

P. O. Box 800-211, Shanghai 201800, China

[3] Center for applied physics and Technology, Peking University, Beijing 100084,

China

[§] Electronic mail: chen_jing@iapcm.ac.cn

[†] Electronic mail: ycheng-45277@hotmail.com

[‡] Electronic mail: zzxu@mail.shcnc.ac.cn

**Abstract:** Wavelength scaling law for the yield of high-order harmonic emission is theoretically examined for excited state atoms which are optically prepared by simultaneously exposing to an extreme ultraviolet pulse at the resonant wavelength and an infrared pulse at a variable wavelength in the range of 0.8μm-2.4μm. Numerical simulations are performed based on the three-dimensional time-dependent Schrödinger equation (3D TDSE) for Ne and H. We confirm that the harmonic yield follows a $\lambda^{-(4-6)}$ scaling with the single fundamental driving laser pulse; whereas for the optically prepared excited state atoms, a $\lambda^{-(2-3)}$ scaling for the harmonic yield is revealed.

High-order harmonic generation (HHG) from gases offers an approach to a table-top source of coherent, tunable, and ultrafast soft-X-ray radiation. The maximum harmonic energy ($E_{max}$) attainable by HHG is determined by the cutoff law $E_{max} = I_p + 3.17 U_p$ [1], where $I_p$ denotes the binding energy of the target atom and $U_p = E_L^2/(4\omega^2)$ the ponderomotive energy ($E_L$, laser electric field strength; $\omega$, laser frequency). A direct consequence of the cutoff law is that the use of a driving laser pulse at a longer wavelength will effectively extend the cutoff of HHG to higher photon energy. However, only recently had high power femtosecond optical parametric amplifier (OPA) with sufficient intensity as a driving source for HHG become commercially available [2-4], which triggered a growing interest in both theoretical and experimental investigations of strong field physics as well as HHG process in intense mid-infrared laser fields [2-8]. Although it has been experimentally proved that HHG with mid-infrared (IR) driving pulses indeed showed extended cutoff photon energies [4, 5]; unfortunately, the wavelength scaling law reported by several theoretical groups [6-8] reveals that HHG yield follows a $\sim \lambda^{-(5-6)}$, which indicates that the HHG efficiency will drop rapidly with the increasing wavelength of driving pulse. This will become a major hurdle for generating intense harmonic radiations in water window or even keV region with IR driving laser pulses.

Currently, several techniques have been established for boosting the HHG efficiency. One of them is phase matching, by which a harmonic field can coherently build up over a long interaction length [9-10]. It has been theoretically demonstrated that phase matching condition can be more easily realized for HHG with an IR driving pulse [11], and this subject certainly deserves more investigation both theoretically and experimentally. Another technique, which can enhance the HHG efficiency based on single atom response to the strong laser field, uses excited gas media prepared by simultaneously exposing atoms to a resonant extreme ultraviolet (EUV) pulse and a near-visible driving pulse [12-14]. The resonant EUV pulse is significantly weaker than the visible pulse, whereas its intensity is sufficiently high to efficiently promote the atoms to their excited state by a single photon excitation process. Subsequently, the excited state atoms can be ionized in the intense visible laser field at much higher rates than the ground state atoms and leads to even more than ten orders of magnitude enhancement of HHG yield [13].

In this Letter, we present a theoretical investigation on the wavelength scaling for HHG in optically prepared excited state atoms aiming at boosting the yield of harmonics generated in the IR driving laser field. A numerical solution of the time-dependent Schrödinger equation (TDSE) within a single-active-electron approximation (SAE) [15] is used for the calculation. The excited state is prepared using an EUV pulse with the photon energy equal to the energy gap between the ground state and the first excited state. The Schrödinger equation of the system is written as

$$i\frac{\partial}{\partial t}\psi(\vec{r},t) = \left[ -\frac{1}{2}\nabla^2 - \frac{Z_{eff}}{r} + \vec{r}\cdot\vec{E}(t) \right]\psi(\vec{r},t), \qquad (1)$$

where $Z_{eff}$ is the effective charge of the nucleus and $\vec{E}$ is the external field

$$E(t) = f(t)(E_0 \sin(\omega t) + E_1 \sin(\omega_1 t)) \qquad (2)$$

with $f(t)$ being the envelop shape of the pulse and $E_0$, $\omega$ and $E_1$, $\omega_1$ being the amplitude and frequency of the IR and EUV fields, respectively. In the calculation, $f(t)$ is chosen to be a flat-top function with two-optical-cycle linear ramps of turn-on and turn-off and a plateau of eight optical cycles. The frequency of the EUV field is equal to the energy gap between the ground state and the first excited state. A model Neon (Ne) atom with adjustment of the effective nuclear charge to fit its ground state energy is used in the calculation of HHG spectra for wavelengths ranging from 800 nm to 2400 nm for the fundamental pump IR field. Figure 1(a) shows the harmonic spectra of model Ne in the IR field at 800nm wavelength with and without the EUV pulse. It can be found that the harmonics in the plateau regime are significantly enhanced by approximately two orders of magnitude when the resonant EUV field is applied and this enhancement can be attributed to the enhancement of the ionization due to the effective excitation of the atom from the ground state to the excited state by the EUV field. This picture is clearly illustrated in Fig. 1(b) where the populations of the ground state are depicted in both cases. The depletion of the ground state has accordingly been enhanced by approximately two orders of magnitude when the EUV field is used, which is consistent with the amount of enhancement in the harmonic spectrum.

It is recently reported that the wavelength scaling law of HHG yield within a fixed energy interval has a large negative index of ~-5--6 [6-8]. Moreover, further calculation with much finer $\lambda$ grid shows rapid oscillations in the yield-wavelength curve. This oscillation with a period depending on the wavelength region is attributed to the interference of up to five different rescattering trajectories [7]; or, from another point of view, the thresholding phenomena within a system of interacting ionization and HHG channels [8]. The wavelength scaling law of HHG yield obtained with the single-color driving field at the fundamental wavelength is well reproduced in our calculation as shown by the solid squares in Fig. 2(a) for the model neon atom; however, as shown by the solid triangles in Fig. 2(a), it is intriguing to note that the index of the wavelength scaling law dramatically decreases to -2 ~ -3 for the fixed energy interval of HHG when the resonant EUV field is applied. For comparison, the results for a hydrogen atom are also presented in Fig. 2(b). The index of the scaling law changes from ~-5 to -3 ~ -4 for hydrogen with the parameters used in our calculation. Such improvement of the scaling law indicates that the use of a combined IR and EUV pump pulse may provide an efficient way toward generation of intense short-wavelength coherent radiation.

It has been pointed out in Ref. [7] that the scaling law ($x \approx -5$) results from the combination of two effects, namely, a factor of $\lambda^{-3}$ due to the spreading of the returning wave packet [16]; and the other factor of $\lambda^{-2}$ owing to the increase of cutoff $E_c \propto \lambda^2$. In order to get the insight into the new scaling law for the combined IR-EUV laser fields, we apply a semiclassical model based on the SFA [16, 17] in which the HHG process is described by a three-step picture. In this model, the time-dependent dipole moment can be expressed as [17]

$$d(t_f) = \sum_{t_i} b_{ion}(t_i) \frac{e^{-iS(t_i,t_f)}}{(t_f - t_i)^{3/2}} C_{rec}(t_f) + c.c., \qquad (3)$$

where $b_{ion}$ is the ionization amplitude at $t_i$, $C_{rec}$ is the spontaneous recombination amplitude at $t_f$ and $e^{-iS(t_i,t_f)}/(t_f - t_i)^{3/2}$ represents the evolution of the electron in the laser field with the classical action defined as

$$S(t_i, t_f) = \frac{1}{2} \int_{t_i}^{t_f} (p + A(t'))^2 dt'. \qquad (4)$$

There is a sum over $t_i$ corresponding to trajectories of electrons that are ionized at different moments but return to the parent ions at the same recombination time $t_f$. However, our calculation shows that the general feature of wavelength scaling law of HHG yield can be faithfully predicted by only considering the shortest two trajectories, as the HHG yield contributed by the multiple-return trajectories obeys the same scaling law as that contributed by the longer trajectory of the shortest two trajectories alone, as shown in Fig. 3. Therefore, for the sake of simplicity, we restrict our discussion only for the shortest two trajectories which are denoted as the short and the long trajectories as follows.

The results of SFA calculation are shown in Fig. 3 for both cases of IR field alone and the combined IR-EUV field [18]. The HHG yields show scaling laws with index $x \approx -4.4$ and -3.4 for the single-color IR field and the combined field, respectively. Furthermore, it is intriguing to note that the HHG yield is dominantly contributed by the long trajectory in the single IR field whereas by the short trajectory in the combined field. In the former case, the short trajectory contribution is negligible in the total HHG yield and decreases much faster with increasing wavelength (with index of the scaling law $|x|>10$) than HHG contributed by the long trajectory. In the latter case of using the combined laser field, on the contrary, the contribution of the long trajectory is negligible and shows a scaling law $x \approx -4.4$ similar to that obtained in the single IR field; while the short trajectory dominates which decreases at a much lower rate of $\lambda^{-3.8}$ with the increasing wavelength, as shown in Fig. 3(b). In addition, in order to reveal the influence of multiple-return trajectory on the scaling law, we present the HHG yield as a function of wavelength for the second-return trajectory, as depicted by the open circles in Fig. 3. Here the second-return trajectory includes the third and forth shortest electron trajectories (corresponding to the $\tau_n$ trajectories denoted as n=3 and 4 in Fig. 3(a) of Ref. [6]). It is found that though the second-return trajectory contributes about equally to the HHG yield as the long trajectory (i.e., the second shortest trajectory), it will not change the overall scaling law in both single IR and IR-EUV fields since it obeys the same scaling law as the long trajectory component. It is worthy to point out here that all the higher-order-return trajectories behave similarly to the second-return trajectory according to our calculation. Therefore the multiple-return trajectories are not included in the following analysis.

For a specific order of high harmonics at a certain wavelength, its efficiency is mainly determined by two aspects according to the well-known HHG mechanism (see Eq.

(3)), namely, the ionization amplitude and the recurrence time of the wave packet which determines the spreading of the wave packet. To gain insight into the wavelength scaling law of HHG yield, we present the ionization amplitudes and recurrence times of both the short- and long-trajectories corresponding to a harmonic radiation with a photon energy of 40 eV upon recombination in Tab. (1) at three different wavelengths $\lambda = 800$, 1200 and 1600 nm. From the Tab. (1), it is easy to find out that for a fixed harmonic energy in the single fundamental IR field, since the ponderomotive energy scales as $\lambda^2$ and the photon energy of the fundamental field scales as $\lambda^{-1}$, the ionization amplitudes and recurrence times of the short and long trajectories have different wavelength scaling laws. As shown in Tab. (1), for the harmonic at the photon energy of 40 eV, the recurrence time of the short trajectory increases much more slowly with $\lambda$ while the ionization amplitude drops by several orders of magnitude when the wavelength increases from 800 nm to 1600 nm. Consequently, the power law of the contribution from the short trajectory is large ($|x| > 10$ as shown in the SFA calculation). On the other hand, the ionization amplitude for the long trajectory changes slowly while the recurrence time increases rapidly with the increasing wavelength which results in a power law considerably higher than 3. In addition, by comparing the ionization amplitudes and recurrence times between the short and long trajectories, one can see that though the diffusion time of the short trajectory is shorter than that of the long trajectory, the ionization amplitude of short trajectory is also much lower ( e. g., $(\Delta t_L / \Delta t_S)^3 \approx 4$ and $(W_L / W_S)^2 \approx 72$ for $\lambda = 800$ nm, as given in Tab. I). Hence the contribution from the long trajectory dominates the HHG yield in a single IR laser field, which is in agreement with the SFA calculation.

In contrast, in the combined IR-EUV field, the high harmonics are mainly contributed by the electron ionized from the first excited state in the over-barrier regime, thus the ionization amplitudes of short and long trajectories will actually have similar values [19]. Therefore the contribution from the short trajectory will *always* dominate in HHG process because of its intrinsically short recurrence time, leading to a new wavelength scaling law with smaller index. This analysis is consistent with the present numerical and SFA calculations.

It is noteworthy that for calculating the results shown by Fig. 2, we assume that the IR and EUV fields have a same pulse duration, a zero phase delay, and a complete temporal overlap. In experiment, this highly coherent superposition is certainly difficult to achieve. Hence we perform more calculations at different time and phase delays between the IR and EUV pulses. The result shows that the relative phase does not affect the scaling law. However, the index of the scaling law in the combined IR-EUV field will increase with the time delay. This can be easily understood since the change of the scaling law in the two-color field is due to the fact that the electron is excited from the ground state to the excited state and then it can be more effectively ionized by the IR field. This process is not dependent on the relative phase between the two pulses. However, when the delay between the IR and the EUV pulses increases, the atoms cannot be efficiently promoted to the excited state, resulting in a reduced enhancement of HHG. In order to achieve an easy synchronization between the IR and EUV pulses, we propose to use an EUV pulse with a pulse duration longer than that of the IR driving pulse. In addition, simulations have also been performed for the model Ne in the single IR field with higher intensity $I = 3.4 \times 10^{14}$ W/cm$^2$ (blue triangle) which gives rise to the same ionization probability as the two-color case (red circle), as shown in Fig. 2(a). One can clearly see that in this case, the HHG yields are similar when the driving wavelength is short. However, the HHG efficiency obtained by use of the combined IR-EUV field is significantly higher than that obtained by use of the single IR field when long driving wavelengths are chosen because of the different wavelength scaling laws. This is similar to the result of hydrogen shown in Fig. 2(b). In this case, the ionization potential of Hydrogen is low and hence the ionization probabilities in both IR field and IR+EUV field are similar. Therefore, the HHG yields in short wavelength regime are comparable but differs significantly in the long wavelength regime due to their different scaling laws.

In summary, numerical evolution of 3D TDSE is performed to investigate the wavelength scaling law for the yield of atomic HHG in which the atoms are optically prepared in the excited state by simultaneously exposing to a combined field of an EUV pulse at the resonant wavelength and an IR pulse at a variable wavelength in the range of 0.8μm-2.4μm. It is found that the HHG yield of specific harmonic in this combined field decreases slower with the increasing wavelength than that in the IR field alone. Our analysis shows that the improvement in the wavelength scaling law in HHG can be mainly attributed to the fact that the atom is pumped to its excited state by the resonant EUV pulse and its ionization mechanism in the IR laser field has been changed from tunneling ionization to over-barrier one. In addition, it is well-known HHG contributed by the short trajectory is usually brighter than that contributed by the long trajectory because of the fact that phase matching condition can be more easily fulfilled for the short trajectory [20]. It can therefore be expected that the use of combined resonant EUV field and long wavelength IR field may provide an effective way to generate intense coherent soft X-ray radiation.

The authors acknowledge the contributions from Yuxin Fu and Yongli Yu at SIOM. The work was supported by the National Basic Research Program of China (Grant No.2006CB806000)，the NNSFC under Grant No. 10574019, CAEP Foundation No. 2006z0202 and No. 2008B0102007. Y. Cheng acknowledges the supports of 100 Talents Program of the Chinese Academy of Sciences and National Outstanding Youth Foundation.

18. In the SFA calculation of the combined field, only the ionization potential of the ground state is replaced by that of the excited state in the ionization amplitude $b_{ion}$ for simplification since most of the photoelectrons are ionized by the IR field from the excited state as shown in Fig. 1(b).

19. Strictly speaking, the ionization amplitude is not applicable here since the ionization amplitude is too large (larger than 1 a.u.). However, this amplitude can be used here to illustrate that there is no significant difference in the ionization process for short and long trajectory components in the IR-EUV field.

**Captions of table and figures:**

Table I: The ionization moment $t_0$, recurrence time $\Delta t$, ionization amplitude from the ground state ($W$) and ionization amplitude from the first excited state ($W'$) for short (with subscript S) and long trajectories (with subscript L) at different wavelengths.

Fig. 1: The harmonic spectra (a) and population of the ground state as a function of time in the IR field without (b) and with (c) the EUV field. The peak intensity of the IR field is $I = 1.5 \times 10^{14}$ W/cm$^2$ and $\omega = 0.057$ a.u. The EUV field $E_1 = \sqrt{0.002} E_0$.

Fig. 2: The wavelength dependence of the integrated HHG yield in the IR field with EUV field (red solid circle) and without the EUV field (black solid square and blue triangle) for Ne (a) （25 eV-50 eV）and H (b) （25 eV-40 eV）. The laser field parameters for Ne are the same of Fig. (1) except $I = 3.4 \times 10^{14}$ W/cm$^2$ for blue triangle points. For H the peak intensity of the IR field is $I = 1.0 \times 10^{14}$ W/cm$^2$. The EUV field $E_1 = \sqrt{0.01} E_0$. The pulse is 40 optical-cycle long totally with five-optical-cycle linear ramps of turn-on and turn-off.

Fig. 3: SFA calculation of the integrated HHG yield （25 eV-40 eV）for Ne initially in the ground state (a) and the first excited state (b). Solid circle: contribution from the short trajectory component; solid triangle: contribution from the long trajectory component; open circle: contribution from the second-return trajectory component; solid square: sum of short, long and second-return trajectories.

Tab. 1.

| λ (nm) | $t_{0S}$ (rad) | $\Delta t_S$ (fs) | $W_S$ (a.u.) | $W_S'$ (a.u.) | $t_{0L}$ (rad) | $\Delta t_L$ (fs) | $W_L$ (a.u.) | $W_L'$ (a.u.) |
|---|---|---|---|---|---|---|---|---|
| 800 | 0.61 | 1.27 | 1.25e-4 | 1.98 | 0.11 | 2.14 | 9.1e-4 | 2.140 |
| 1200 | 0.84 | 1.41 | 9.1e-6 | 1.70 | 0.04 | 3.54 | 9.5e-4 | 2.141 |
| 1600 | 0.96 | 1.58 | 8.7e-7 | 1.45 | 0.02 | 4.90 | 9.6e-4 | 2.142 |

Fig. 1

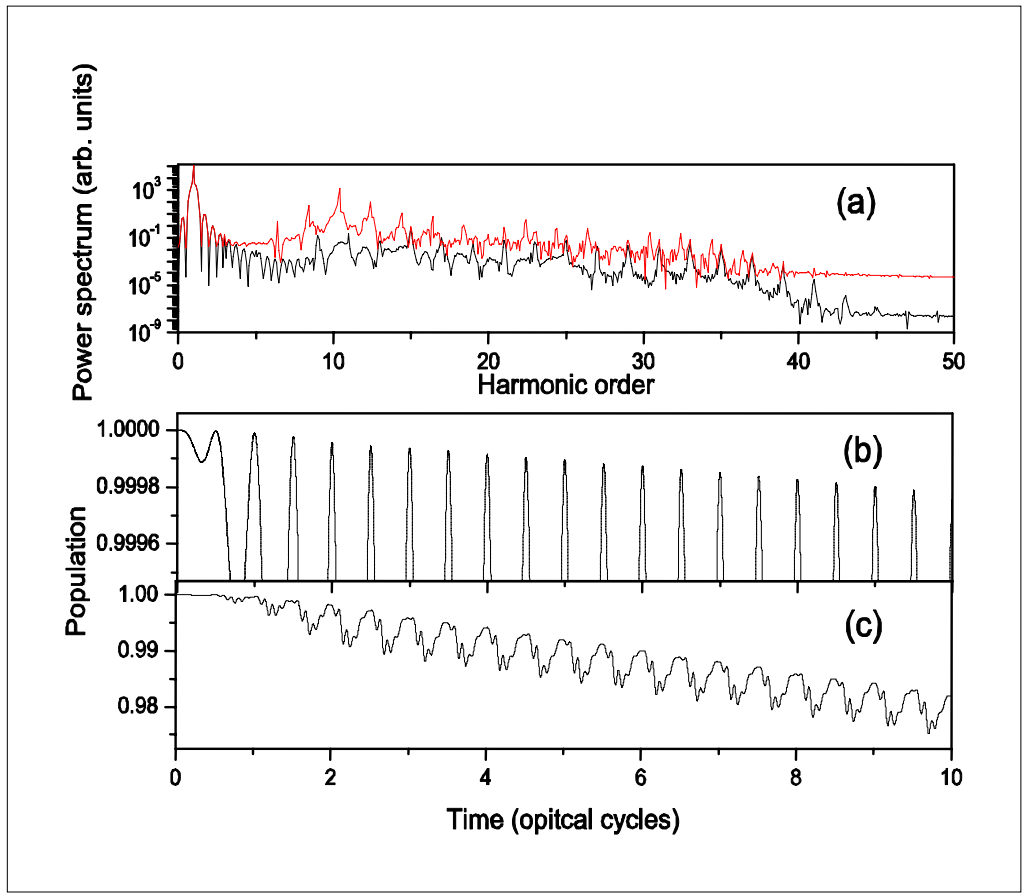

Fig. 2

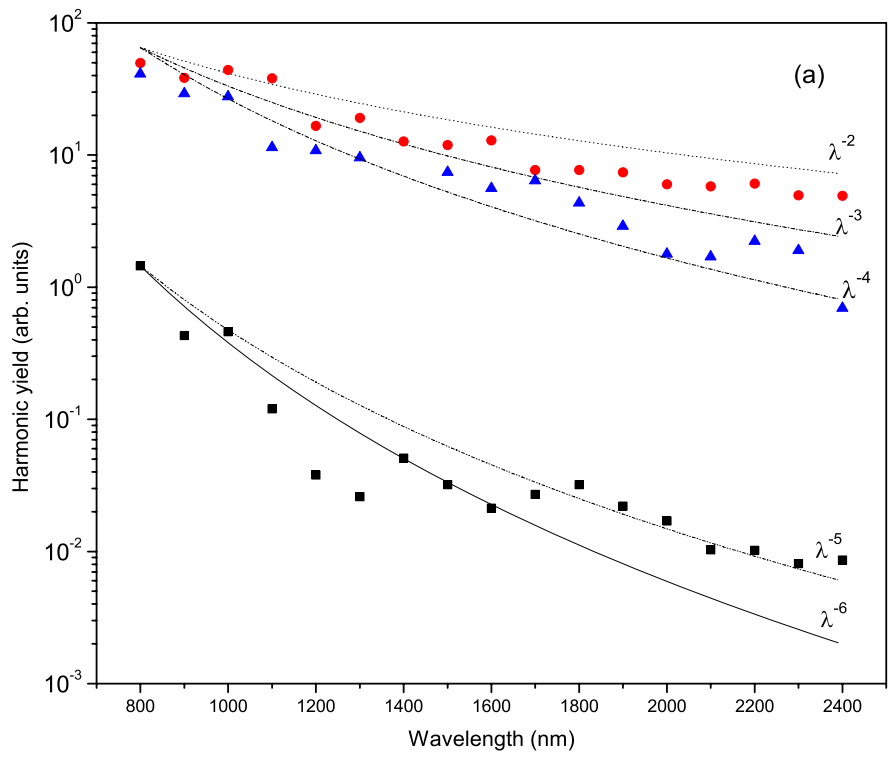

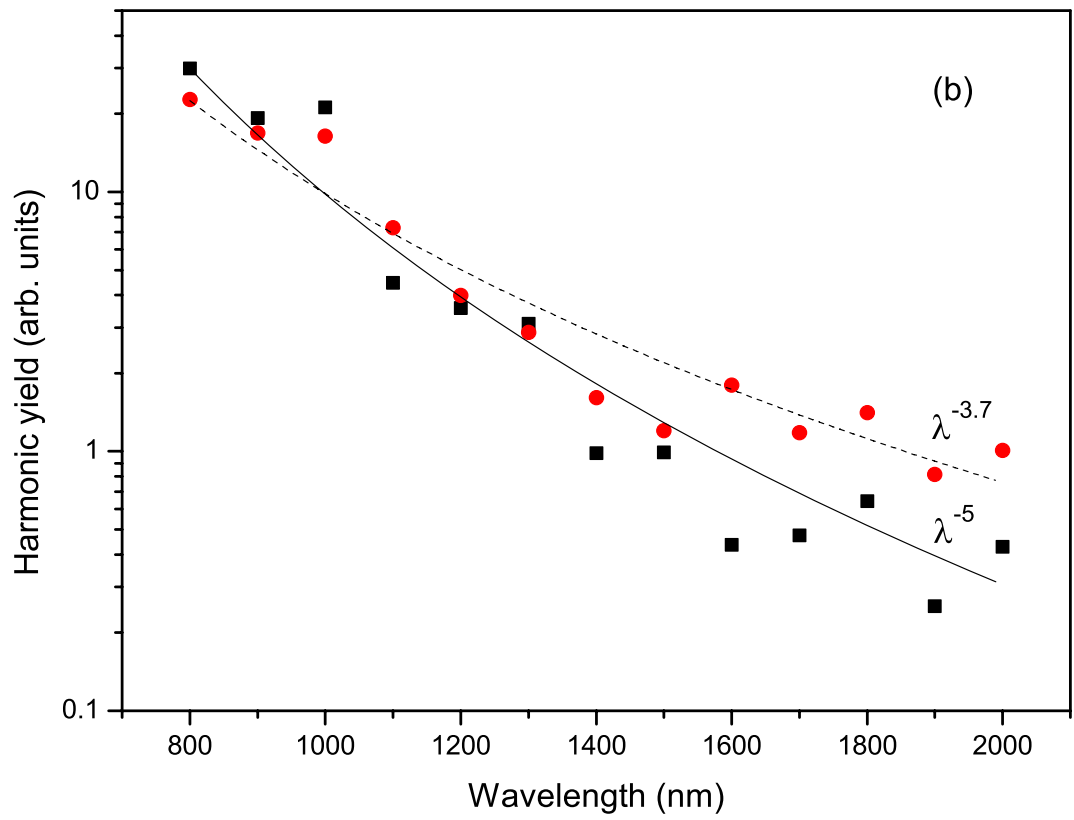

Fig. 3

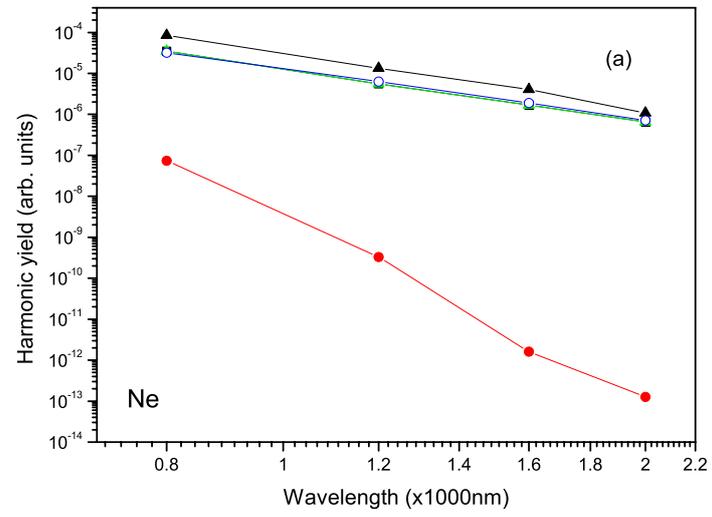

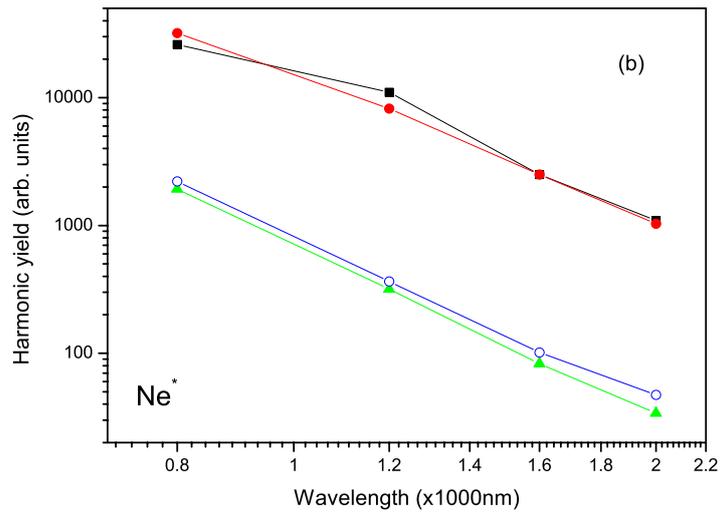